\documentclass[11pt,twoside]{article}

\usepackage{asp2006}
\usepackage{epsf}
\usepackage{lscape}
\usepackage{graphicx}

\begin{document}

\title{AGB stars and the chemical evolution of galaxies}   
\author{Monica Tosi}   
\affil{INAF - Osservatorio Astronomico di Bologna, Via Ranzani 1, I-40127
Bologna, Italy}  

\begin{abstract} 
Asymptotic Giant Branch (AGB) stars are important players in the 
chemical evolution modelling of galaxies, because they are major producers 
of several chemical elements and excellent tracers of the structure and of 
the star formation activity
of their parent galaxies. A few examples on the importance of AGB
stars are presented in this review, together with a number of open problems 
affecting chemical evolution  model predictions related to the element 
enrichment by AGB stars: the evolution of $^4$He and Na in
globular clusters, the evolution of $^3$He and the carbon isotopes in the 
Galactic disk, and the evolution of N and O in different types of galaxies. The
need of homogeneous and complete sets of yields is emphasized.
\end{abstract}


\section{Introduction}   
There are several reasons why galaxies and galaxy modellers care about 
Asymptotic Giant Branch stars. AGB stars
trace the halo stellar distribution (see Demers in 
this volume), thus indicating what is the size and the dynamics of the system; 
they trace intermediate-age stars (see Grebel, this volume),   
 thus showing if and how much star formation (SF) activity has occurred in the
 region 0.5 - 1 Gyr ago; they are the major site of production of 
 several chemical elements (see Busso and Lattanzio in this volume) and
fundamental contributors to the stellar nucleosynthesis yields, which, in turn,
are the main ingredients of chemical evolution models.

This review is focussed on the third point, which is the most important one  for
chemo-dynamical models of galaxy evolution. Let me however mention how 
important it is to trace intermediate age and old populations. 
IZw18 is the most metal poor star-forming
galaxy ever discovered. Being very blue and full of gas, it was
often considered a local counterpart of primeval galaxies and
 a really young system, with SF activity begun only a few
Myr ago. However, when HST optical photometry became available,  AGB stars
were found \citep*{Al99} and then confirmed 
with near-infrared HST photometry \citep{O00}. 
The presence of AGB stars undisputably demonstrated
that IZw18 is not as young as originally thought and must
have started forming stars at least 0.5~-~1 Gyr ago. 

Here, I will describe the effects of AGB 
stars as interstellar medium (ISM) polluters
with reference to the evolution of a few interesting elements: $^3$He, $^4$He,
$^{12}$C, $^{13}$C, N and Na. 
The galaxy evolution models described here are all {\it standard}, in the 
sense that they do not explicitely treat dynamical aspects. It is however 
becoming increasingly clear that to understand galaxy evolution dynamical and 
hydrodynamical effects cannot be left aside. 

\section{AGB stars as ISM polluters: $^4$He, Na and the evolution of globular clusters}

In recent years there has been an increasing interest on the possibility of 
a second generation of stars in some globular clusters. Some of the 
observed properties of globular clusters are indeed considered evidence 
 of a second SF event. For instance, the 
anticorrelation  between the sodium and oxygen abundances measured in stars 
of several clusters \citep*[e.g.][and references therein]{C04} has 
been interpreted as the consequence of the cluster self-enrichment, if the gas
replenishment from the retention of stellar ejecta is sufficient to allow 
for further SF. In this scenario, the stars born during the second SF episode
form from mostly (if not entirely) recycled gas and their initial chemical 
compositions reflect the yields of the stars mostly contributing to the cluster
self-enrichment. \citet{G01} and \citet{D02} suggested that
high Na enrichment and significant O depletion are most naturally explained if
the major culprits of the cluster self-pollution are relatively massive AGB 
stars. The stars with more O and less Na would be those formed 
in the first generation, while the stars with less O and more Na would be 
those formed in the second generation.
\citet{D02} further suggested that the second generation would be
significantly enriched also in helium (again as a consequence of the
predominance of AGB star ejecta in the gas available for SF) and that this 
could explain the Horizontal Branch morphology of clusters with extreme blue 
tails. Other authors, however, have argued that the winds of massive stars have
more chances than AGB stars to adequately pollute the cluster's medium without
the side effects of requiring unusual initial mass functions and 
stellar remnants \citep[e.g.][]{PC06}. Appropriate chemo-dynamical
models are required to test advantages and disadvantages  of the various 
hypotheses.

The discovery of a second, bluer  Main Sequence (MS) in $\omega$ Centauri 
\citep{B04}, together with that  of multiple subgiant and red giant branch (RGB)
 sequences, 
also calls for the existence of multiple stellar generations, with the
additional striking surprise, provided by high-resolution spectroscopy, that the 
bluer MS is 0.3 dex less metal poor than the {\it standard} red MS \citep{P05}. 
Piotto et al. argue that the only way to allow for the measured 
colour shift between the two MSs with the measured metallicity difference is 
to let the bluer MS be much more helium rich than the other, with a difference
in the helium mass fraction $\Delta$Y=0.14. 
If we attribute to the red MS a primordial He mass fraction of
Y=0.24, this implies that the blue MS should have Y=0.38: an abundance higher
than in any other observed star cluster or galaxy !

Would AGB stars be able to provide such a huge helium enhancement ? It is very
unlikely. However, $\omega$ Centauri is definitely not a normal globular
cluster; may be not a cluster at all, but the remnant of a nucleated dwarf 
galaxy captured
and stripped by the Milky Way, with the current cluster actually being the
original nucleus of the satellite. 
\citet{BN06} have recently shown that
the observed properties of $\omega$ Centauri cannot be explained without
considering the strong dynamical interactions with the Galaxy. The actual
question then is whether the medium, out of which subsequent stellar generations have
formed, was enriched by $\omega$ Centauri own stars or by the stars of the host
dwarf originally surrounding it. Whether the polluters were AGBs, massive stars
or supernovae is in this case a second level issue.

\begin{figure}[!ht]
\begin{center}
\includegraphics[width = 388pt, height = 200pt]{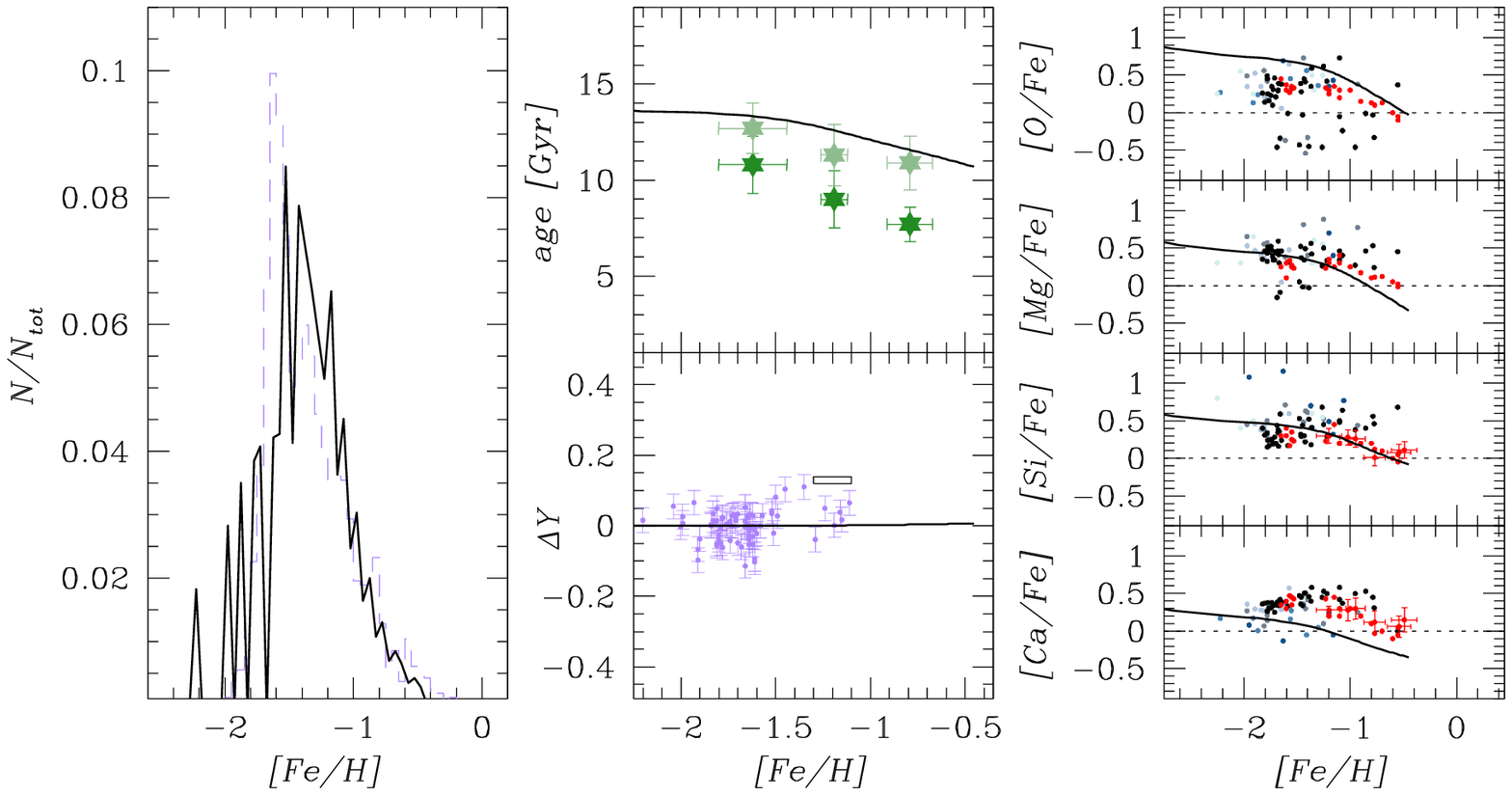}
\end{center}
\caption{Predictions from chemical evolution models (solid lines) for 
$\omega$ Centauri compared with the 
corresponding observational data (dots and dashed lines). Left panel: the
stellar metallicity distribution from \citet{S05}; central top panel:
age-metallicity relation from \citet{H04}; central bottom panel: helium vs iron
from \citet[][box]{P05} and in RR Lyraes \citep[dots; from][]{S06}; 
right panel: abundance ratios vs iron from several
literature sources. }
\label{omega}
\end{figure}

We are currently modelling the chemical evolution of $\omega$ Centauri (Romano
et al., in preparation) considering it as a dwarf galaxy and following the
approach presented by \citet*[][hereafter, RTM06]{RTM06} for other dwarfs. 
We adopt the SF history derived from the \citet{S05} data and  
allow for both galactic winds and infall of metal poor gas. The model
predictions are compared with the observational mass, age-metallicity
relation, metallicity distributions, chemical abundances and abundance ratios.
In agreement with Bekki \& Norris results, in no way are we able to 
obtain model predictions consistent with the data if we consider 
$\omega$~Centauri as an isolated system. 
On the other hand, by considering it as the residual of a nucleated galaxy
stripped by the Milky Way 10 Gyr ago, we reproduce rather well all the data,
except the extremely high He abundance of the blue MS. To achieve this goal, new
ad hoc assumptions are needed and will be the subject of further efforts,
involving appropriate assumptions on the fate of the stellar ejecta and
new stellar yields for both high and intermediate mass stars.
Our current, preliminary results are shown in Fig.\ref{omega}.

\section{AGB stars as ISM polluters: $^3$He, $^{12}$C and $^{13}$C }

$\omega$ Centauri may require special yields and evolutionary conditions to be
explained, but the need of improved yields is a much more general problem. 
Examples of important elements for which
improved yields are badly needed are the He, C, N and O stable isotopes. 

In the late nineties it has been shown 
\citep[e.g.][and references therein]{G97,T00} that to let 
the predictions of Galactic chemical evolution models reproduce the low 
$^3$He abundances measured in Galactic HII regions a mechanism is needed, able 
to drastically reduce the $^3$He production normally predicted for low-mass 
stars. Such a mechanism was suggested by e.g. \citet{C95} and \citet{W95} to 
be the consequence of extra-mixing at work in RGB stars, possibly as a 
consequence of rotation.  To reconcile the low HII regions abundances with
the high $^3$He measured in a few Planetary Nebulae (PNe), the extra-mixing
should affect about 90\% of low mass stars. Since this extra-mixing implies not
only a significant $^3$He depletion, but also a larger conversion of $^{12}$C into 
$^{13}$C in such a large fraction of stars, it is important to check whether 
the corresponding yields are consistent with the carbon ratios observed in PNe. 

\begin{figure}[!ht]
\centerline{
\scalebox{1.0}{%
\includegraphics[width = 180pt, height = 200pt]{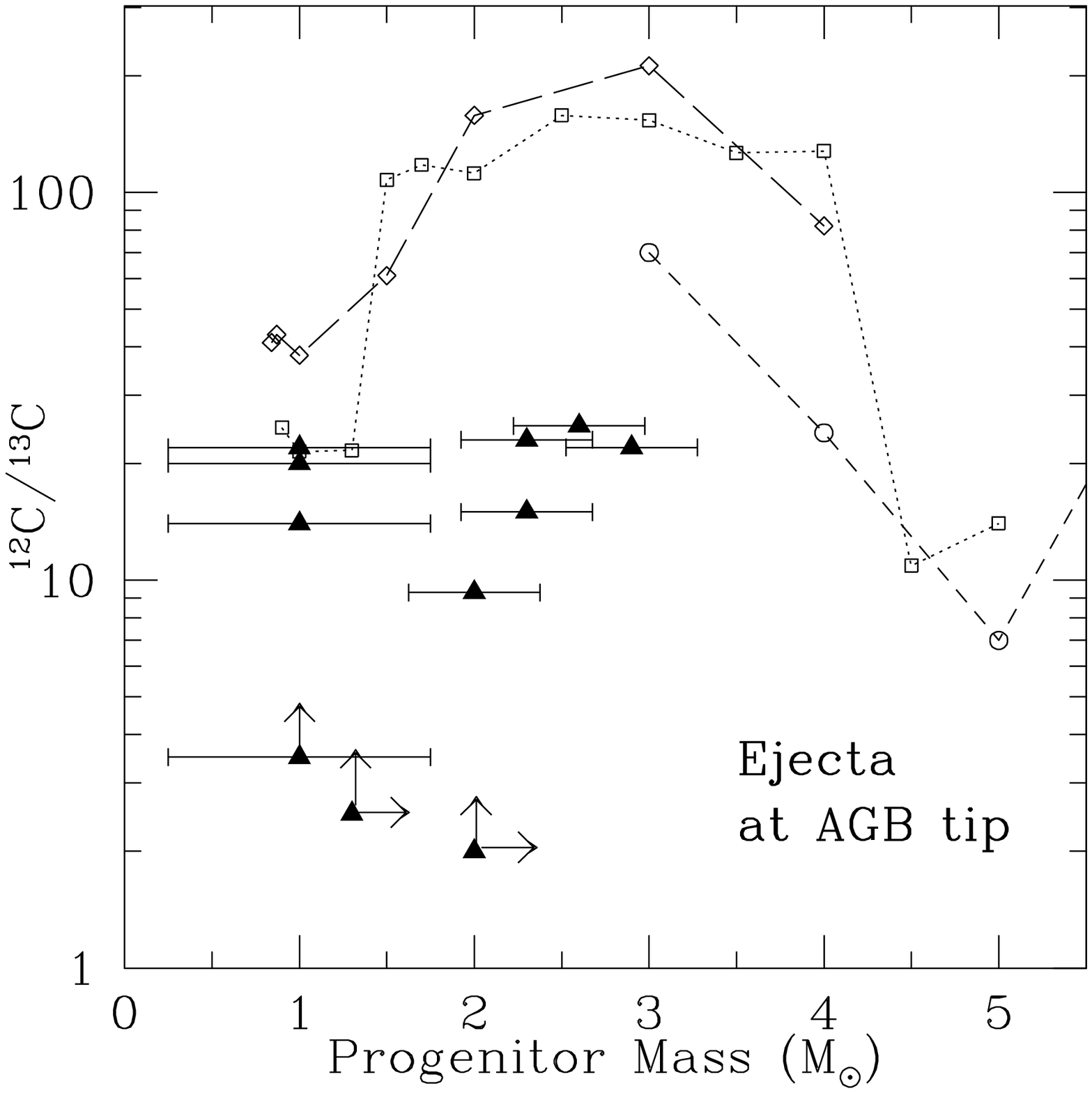}}
\scalebox{1.0}{%
\includegraphics[width = 180pt, height = 200pt]{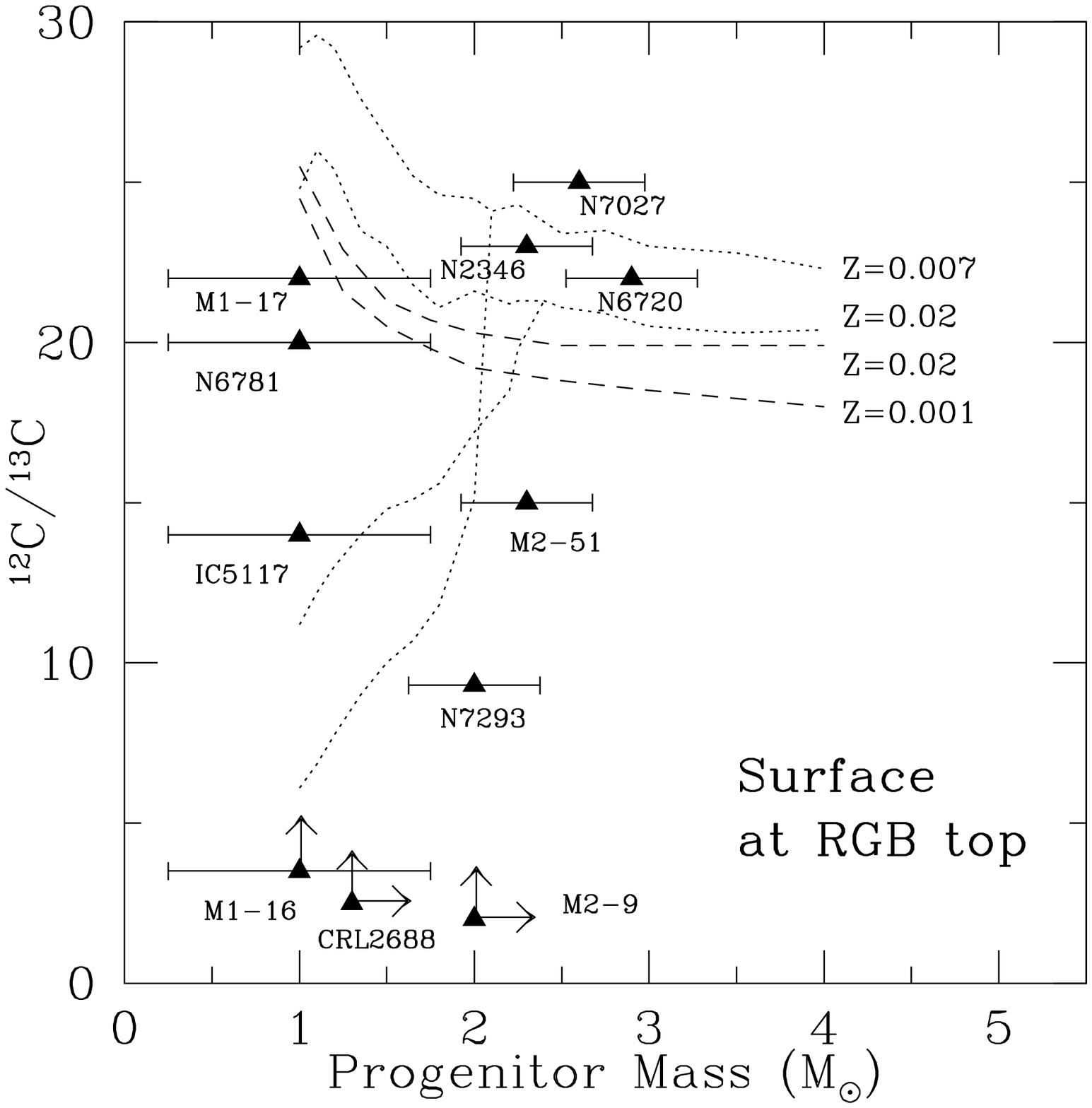}%
}
}
\caption{From \citet{P00}: carbon isotopic ratios measured in PNe (dots with 
error bars in both panels) compared with stellar nucleosynthesis predictions.
The curves in the left-hand panel show the ratio predicted without deep-mixing
just before the PN ejection by \citet{M01}, long-dashed, \citet{vdG97}, dotted,
and \citet{FC97}, dashed. The dotted curves in the right-hand panel show the 
ratio predicted at the end of the RGB phase by \citet{BS99}, with (the curves
falling down to low ratios) and without deep-mixing (those staying at high
ratios).  }
\label{carbon}
\end{figure}

In Fig.\ref{carbon} the carbon isotopic ratios measured by \citet{P00} in 
several PNe are compared with the predictions from various theoretical yields.
It is apparent that the standard nucleosynthesis predictions of the left-hand
panels overproduce the carbon ratio, while the deep-mixing predictions by
\citet{BS99} in the
right-hand panel nicely fit the data. This nice fit is however misleading, 
because Boothroyd \& Sackman's computations reach only the RGB tip and not the final
evolutionary phases. The carbon abundances can be significantly affected 
by the remaining evolution. Unfortunately, no yields taking deep-mixing into
account have been calculated beyond the RGB and none of the yields computed up
to the final phases include the deep-mixing effect. In other words, a direct
check of the existence of the extra-mixing and of the solution to the $^3$He
problem, with all its cosmological consequences, is currently unfeasible !

\section{AGB stars as ISM polluters: nitrogen versus oxygen}

\begin{figure}[!ht]
\centerline{
\scalebox{1.0}{%
\includegraphics[width = 188pt, height = 246pt]{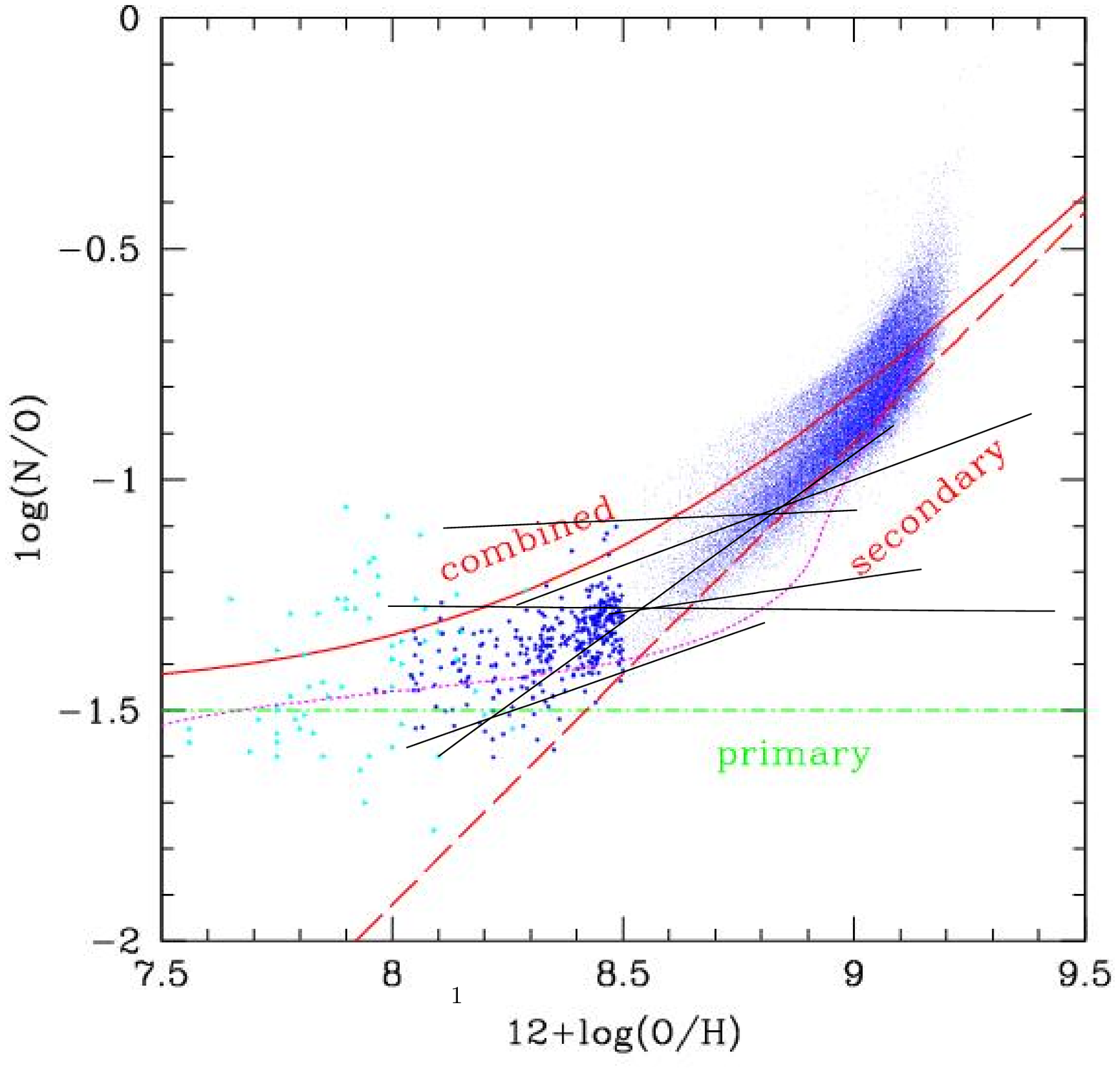}}
\scalebox{1.0}{%
\includegraphics[width = 188pt, height = 222pt]{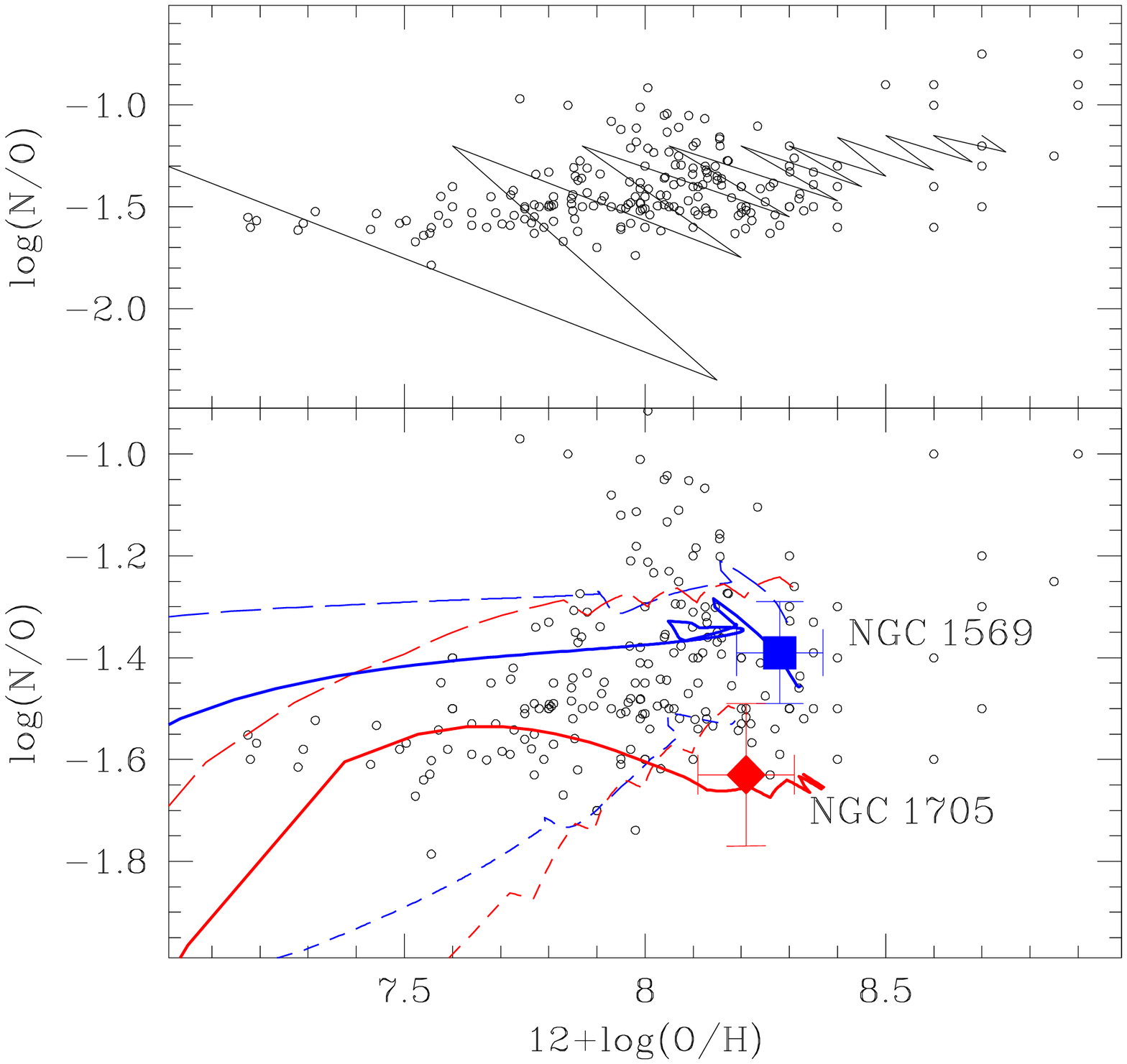}%
}
}
\caption{N/O vs O/H in different galaxies as derived from HII region
observations (dots). The plot on the left-hand panel is from \citet{L06} and
shows the SDSS data as dots. The thick straight lines show the trends
in individual spirals from \citet{DT86}. The right-hand panels are blow-ups of
the metal-poor region of the diagram and the lines show some of the 
effects of galaxy evolution (see text for details).
}
\label{nitrogen}
\end{figure}

Nitrogen and oxygen are important tracers of the chemical evolution of galaxies.
Not only are they among the most abundant elements; they are also measurable, 
from the emission lines of HII regions (and PNe in some cases), in galaxies 
up to rather large distances, where they are often the 
only available metallicity indicators. Oxygen is mainly produced by massive
stars, whilst nitrogen is mostly synthesized by intermediate-mass stars. 
In the framework of the simple model of galaxy evolution
\citep{T80} the enrichments of primary elements are independent of each 
other, whilst the growth of a secondary element goes as the square of the
growth of a primary.  The double nature (primary and secondary) of the N 
production became apparent when the N and O measured in HII regions of a number
of spirals \citep{DT86}, including the Milky Way, and of late-type dwarf 
galaxies \citep{MT85} were compared with the predictions
of adequate chemical evolution models: a significant fraction
(between 30\% and 60\%) of N must be primary to explain the rather flat trend
of N/O vs O/H inferred from HII regions in both late-type dwarfs and 
individual spirals. The straight lines in the left-hand panel of 
Fig.\ref{nitrogen} show least-squares fits to the N/O vs oxygen
abundances derived by \citet{DT86} from the HII regions in the Galaxy, M31, M33,
M101, NGC2403 and IC342 \citep*[for more accurate and updated data see][]{PTV03}.
The fits for these individual spirals are overplotted on
a very recent version of the N/O vs O/H diagram including results from the Sloan
Digital Sky Survey \citep{L06}. 

The N/O vs O/H diagram is sometimes interpreted as an
evolution plot. However, the oxygen abundance in abscissa cannot
be taken as a proxy for time. In fact, each of the plotted dots corresponds to 
an HII region, whose abundance reflects the final (current) result of
the whole history of chemical enrichment in the host galaxy region. Such history is
known to be quite different for different types of galaxies, since it depends on
SF history, gas accretion (infall) and loss (winds), interactions with
companions, etc.  As such, the
position of each point in the plot depends on the different evolutions of
different galaxies, and not only on the nature (primary or secondary) of the N 
nucleosynthesis \citep[see also][]{EP78,PTV03}.  

Much attention has been payed to the metal-poor end of the N/O vs O/H diagram,
where late-type metal-poor dwarfs are located, to discuss either the possibility of 
primary production of N in massive stars, or the possible connection between 
late-type dwarfs and Damped Lyman-$\alpha$ systems. The plots on the right-hand
side of Fig.\ref{nitrogen} show some of the effects on the N and O abundances
occurring during galaxy evolution. The lines shown in the two diagrams on the
right correspond to the predictions of different chemical evolution models for
the time behaviour of oxygen and nitrogen in a late-type dwarf 
galaxy. The top-right panel compares with the abundances measured from the 
HII regions observed in late-type dwarfs (dots) the model predictions by 
\citet{P93} for a generic  Blue Compact galaxy experiencing about ten SF bursts. 
It shows how the HII region self-enrichment and the 
different timescales for the stellar ejection and subsequent diffusion in the 
ISM of oxygen and nitrogen lead to a saw-tooth shape for the N/O ratio as a 
function of oxygen. 

The bottom-right panel show instead the predictions by
RTM06 for the starburst dwarf NGC~1569 (always the line with higher N/O
for each line-type; blue in the colour version of the figure) and the 
Blue Compact dwarf NGC~1705 (always the line with lower N/O
for each line-type; red in the colour version of the figure). The big symbols
with
error bar indicate the corresponding values derived from HII regions in NGC~1569
(blue square) and NGC~1705 (red diamond). The other points
correspond to HII regions values in other late-type dwarfs. Different 
line-types correspond to different assumptions on the stellar yields or on the 
galactic wind efficiency. See RTM06 for references and details.
All the lines show model
predictions for the evolution of oxygen and nitrogen with time and it is
apparent how much the time behaviour can be different from a simple fit to the
current N/O vs O/H distribution inferred from HII regions. 

In spite of the different
classification, NGC~1569 and NGC~1705 have quite similar properties 
\citep[see e.g.,][and references therein]{An03,An05}: similar gas and star 
masses, metallicity, IMF, SF history with very strong recent activity, and
observational evidence of similarly strong galactic winds. Yet,
they have quite different N/O ratios, which require different evolutionary
assumptions. The long-dashed lines in the bottom-right panel 
of Fig.\ref{nitrogen} show the model predictions based on the standard yields by
\citet{vdG97} for intermediate mass stars. Clearly a standard N production  
may be consistent with the N/O ratio observed in NGC~1569, but un-reconcilable 
with the low N/O of NGC~1705.  If we consider yields with smaller N
production, even in the extreme case of those by \citet{MM02} where the
hot-bottom burning phase has not been computed, the models overpredict the N/O
ratio observed in NGC~1705 (short-dashed lines). A better fit to the observed 
abundances is obtained (RTM06) if not only the
nitrogen production in intermediate mass stars is relatively low, as predicted 
for instance by the {\it minimal} hot-bottom burning case proposed by 
\citet{vdG97}, but also a higher efficiency of nitrogen loss in the galactic wind is
allowed in NGC~1705 than in NGC~1569 (solid lines).

Detailed chemical evolution models of individual dwarfs, such as those shown 
here for NGC~1569 and NGC~1705, have become possible only
recently, mostly thanks to HST, which has
allowed to derive their SF histories back to quite early epochs. Two kinds of
models for individual galaxies can be computed: {\it standard} chemical 
evolution models \citep[e.g.][]{C99,LM03,RTM06}  and {\it chemo-dynamical} models
\citep[e.g.][]{Re02,Re06}. The former have the drawback of a simplistic 
treatment of star and SN feedbacks and gas motions, the latter have the problem 
that the timescales appropriate for hydrodynamics make it prohibitive, in 
terms of CPU time, to follow the system evolution over more than 1 Gyr. The
challenge in the next few years is to improve both types of approaches and get a
more realistic insight of how stars and gas evolve, chemically and dynamically,
in their host galaxies.

\section{The need for improved yields}

In the previous sections, some examples of the uncertainties affecting the 
predictions of chemical evolution models have been presented. Part of these
uncertainties are due to the lack of complete and homogenous sets of stellar
yields for various initial metallicities. To compute adequate chemical 
evolution models of whatever galaxy, we need
homogeneous chemical yields for all the major isotopes, for the whole
range of stellar masses, for many initial compositions and taking into account
all the most relevant processes occurring in the stellar interiors until the
final evolutionary phases. Such {\it optimal} grid of yields is far from
existing.  The results from stellar
nucleosynthesis are steadily improving with time, with most of the processes
occurring during stellar evolution being treated with increasing precision.
However, the yields available to the community are still very heterogeneous and
incomplete and not one single set of nucleosynthesis predictions exists taking
properly into account all the processes in all the phases of stars of all 
masses (say from 0.8 to 100 M$_{\odot}$) and at least 2-3 different 
metallicities (from metal poor to
 solar and, possibly, super solar). This circumstance
not only prevents the computation of detailed self-consistent chemical evolution
models for a number of key elements, but can even lead to misleading results.
The potential risk can be visualised by comparing with each other the yields 
provided by different authors and the corresponding normalizations through an
IMF. 

\begin{figure}[!ht]
\begin{center}
\includegraphics[scale = 0.65, angle = 0]{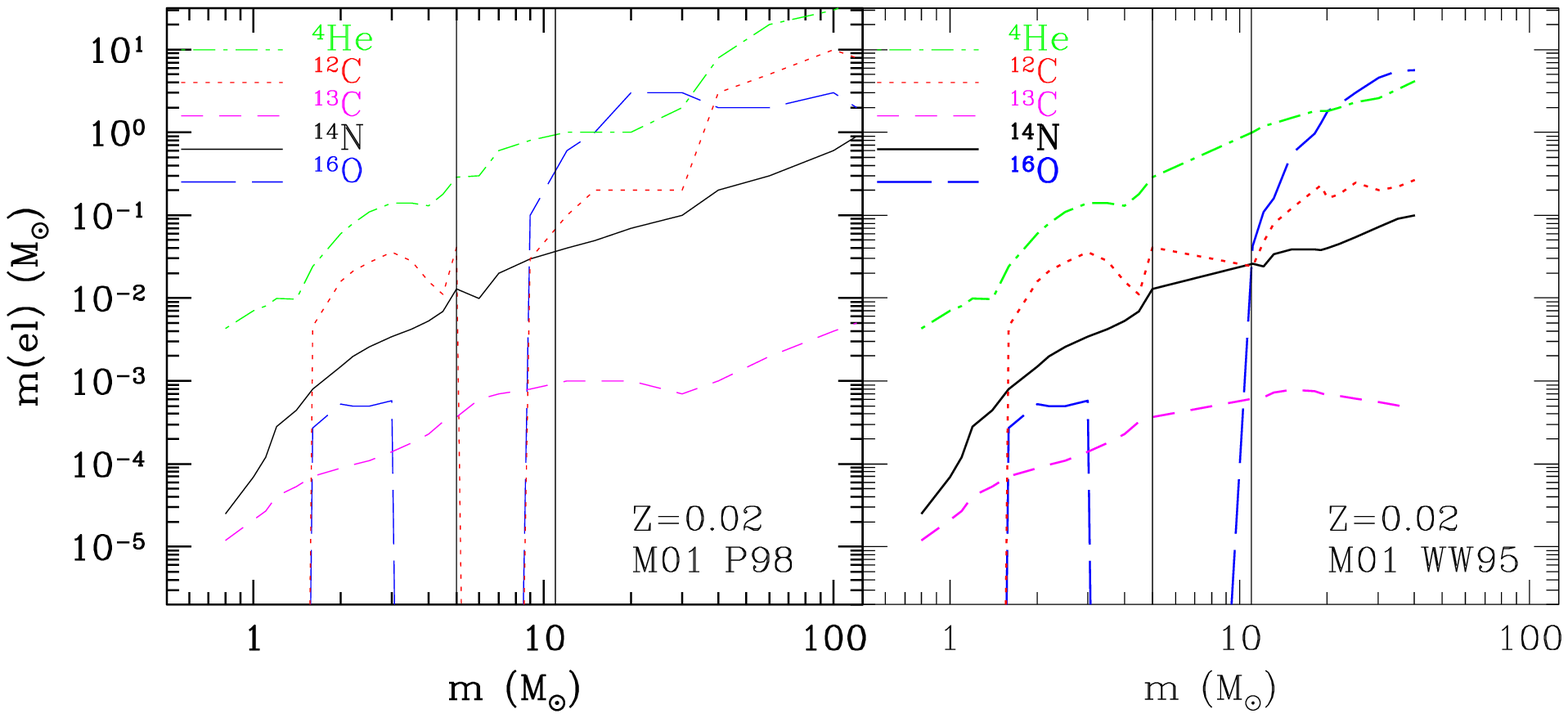}
\end{center}
\caption{Stellar yields: ejected mass of newly synthesized element as a function of the stellar
initial mass. Left-hand panel: the case of the homogeneous sets of solar 
yields by \citet{M01} and \citet{P98} covering the whole mass range. 
Right-hand panel: the {\it standard} case of incomplete coverage with 
inhomogeneous sets \citep{M01,WW95}  which do not consider initial
masses with 5 $<$ M/M$_{\odot} <$ 11 (the lines plotted in this mass range are
linear interpolations) and beyond 40 M$_{\odot}$.
}
\label{yields}
\end{figure}

\begin{figure}[!ht]
\begin{center}
\includegraphics[scale = 0.65, angle = 0]{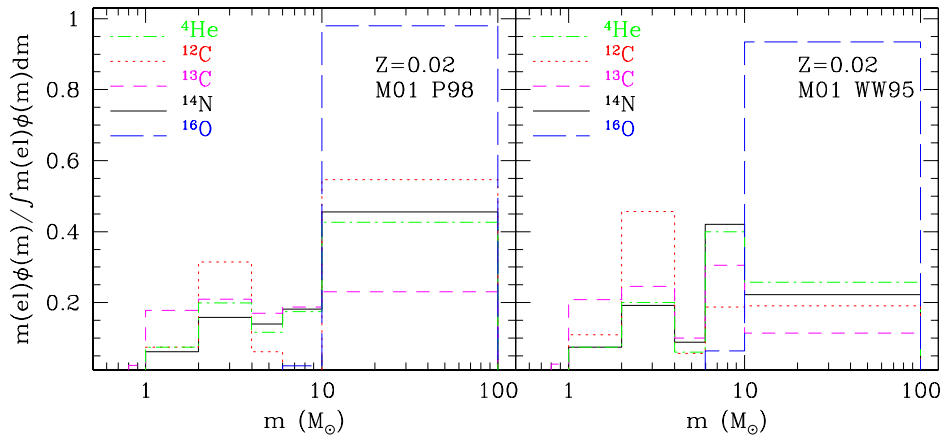}
\end{center}
\caption{Fractional ejected mass of newly synthesized element weighted with
\citet{T80} IMF. Left-hand panel: the case of the homogeneous sets of solar 
yields by \citet{M01} and \citet{P98}. Right-hand panel: the case of the 
inhomogeneous, non adjacent sets by \citet{M01} and \citet{WW95}.
}
\label{yieldsimf}
\end{figure}

The left-hand panel of Fig.\ref{yields} shows a subset of
the solar yields presented by \citet{P98} for
massive and quasi-massive stars and by \citet{M01} for low and intermediate mass
stars. This is the best case in literature of self-consistent yields for all
masses, based on the same stellar evolution models and input physics. What
is normally available in the literature is  a collection of yields for
partial mass ranges, each computed under different assumptions and often for
different metallicities. A typical case is shown in the right-hand panel of
Fig.\ref{yields}, where the
solar yields by \citet{M01} for low and intermediate mass stars are now combined with
those by \citet{WW95} for massive stars. Two problems are immediately apparent.
First, stars in the mass range 5 $<$ M/M$_{\odot} <$ 11  have not been
computed by either Marigo or Woosley \& Weaver, which implies that to use this combination of yields one
must interpolate over this mass interval. Second, massive stars do not go beyond
40 M$_{\odot}$ and to higher masses one must therefore extrapolate. The latter
issue might not have overwhelming consequences in the modelling of the recent
chemical evolution of the Milky Way, since any reasonable IMF predicts very 
few stars more massive than 40 M$_{\odot}$, but can be extremely relevant 
for very early epochs, when the most massive stars were the only polluters. 
The former problem has very serious implications because stars in the 5 -- 11
M$_{\odot}$ range are the most effective contributors to the ISM
chemical enrichment.
In Fig.\ref{yieldsimf} the yields of Fig.\ref{yields} have been weighted with
Tinsley's (1980) IMF. The linear interpolation performed to cover the
5 -- 11 M$_{\odot}$ interval absent in the Marigo/Woosley\&Weaver combination
results in a bump (right-hand panel) in the contribution of these stars 
to the enrichment of He, N and O which is totally absent in the left-hand panel,
where the homogeneous sets of yields are shown. This enhanced contribution is
most likely spurious and can lead to a significant overprediction of the 
elements mostly produced by stars of these masses.

To overcome these problems, we strongly encourage the community of stellar nucleosynthesis experts
to provide homogeneous yields for all stellar masses, computed up to the final
evolutionary phases and for several initial metallicities.

\acknowledgements 
Some of the results described here have been obtained thanks to pleasant and
recurrent collaborations with A. Aloisi, D. Galli, F. Matteucci, F. Palla, 
D. Romano and L. Stanghellini. I am grateful to Corinne Charbonnel for many
useful conversations on the stellar yields and to Donatella Romano for her 
invaluable help. Part of these researches was funded through INAF-PRIN-2005.






\end{document}